\documentclass[aps,twocolumn,prc,showpacs,preprintnumbers,nofootinbib,float,superscriptaddress,longbibliography]{revtex4-1}

\usepackage{epsfig}
\usepackage{color}
\usepackage[dvipsnames]{xcolor}
\usepackage{float,amsmath,amssymb,diagbox}
\usepackage{graphicx}
\usepackage{url}
\usepackage{bbold}
\usepackage{ulem}
\usepackage[utf8]{inputenc}
\usepackage{hyperref}
\usepackage{lineno}

\usepackage[utf8]{inputenc}

\begin{document}
\title{Search for Dark Photons in Polarized  $\boldsymbol{\gamma\gamma\rightarrow e^+e^-}$ at RHIC}

\author{\it Isabel Xu}
\email{isaxu12193@gmail.com}
\author{\it Nicole Lewis}
\email{nlewis@bnl.gov}
\address {Physics Department, Brookhaven National Laboratory, Upton, NY 11973, USA}
\author{\it Xiaofeng Wang}
\email{xiaofeng_wang@mail.sdu.edu.cn}
\address{Key Laboratory of Particle Physics and Particle Irradiation (MoE), Institute of Frontier and Interdisciplinary Science, Shandong University, Qingdao, China}
\author{\it James Daniel Brandenburg}
\email{brandenburg.89@osu.edu}
\address {Physics Department, Brookhaven National Laboratory, Upton, NY 11973, USA}
\author{\it Lijuan Ruan}
\email{ruan@bnl.gov}
\address {Physics Department, Brookhaven National Laboratory, Upton, NY 11973, USA}

\date{\today}

\begin{abstract}
The fundamental nature of Dark Matter remains one of the major mysteries of modern physics. Some models postulate the existence of a massive gauge boson, a ``dark photon'' ($A^\prime$), that may allow Dark Matter particles to interact with Standard Model particles. Ultra-relativistic heavy-ion collisions produce highly Lorentz-contracted electromagnetic fields with sufficient energy density to potentially manifest as light dark photons. We report limits on dark photon parameters via a search for anomalous production of $e^+e^-$ pairs in $\gamma A^\prime$ fusion from ultra-peripheral $\rm{Au}+\rm{Au}$ collisions.  This study utilizes measurements of the Breit-Wheeler process ($\gamma\gamma \rightarrow e^+e^-$) carried out by STAR and the dark photon as a mediator, specifically making use of polarization-dependent final-state azimuthal asymmetries in $e^+e^-$ pairs. These limits are informative for future searches in the ultra-peripheral heavy-ion collisions and to constrain future theoretical developments of the dark photon mechanism.

\end{abstract}
\maketitle

\section{Introduction}
Over the past several decades, an overwhelming amount of evidence has been uncovered indicating that the majority of the gravitational mass in the Universe is not visible, assuming that general relativity is a complete description of gravity at galactic scales. The anomalous rotational velocities observed within some galaxies is the most commonly cited evidence for this non-luminous mass, indicating that the visible mass is insufficient to explain the gravitational mass within the galaxy~\cite{Zwicky:1933gu,Rubin:1980zd}. Within the standard cosmological model, these observations have been understood in terms of the presence of an unseen gravitational mass, known as Dark Matter (DM). In recent decades, additional evidence based on gravitational lensing effects~\cite{Clowe:2006eq}, and the cosmic microwave background~\cite{Planck:2015mrs} have added additional credibility to this Dark Matter hypothesis. This body of evidence is, to this day, the clearest indication of physics beyond the Standard Model (SM).  Despite the overwhelming evidence for its existence, much  is still unknown about DM, such as how it is created, its composition, and whether it interacts with SM particles in any way other than gravitationally.  
 
Since Dark Matter is expected to be uncharged (or weakly coupled) under all SM interactions, it is exceedingly difficult to detect. Despite lacking direct interaction with SM particles, some models favor self-interacting DM to explain the low-density galactic-scale core distribution of gravitational mass which has been inferred from lensing measurements~\cite{Moore:1994yx}.  Within such models~\cite{Okun:1982xi,Georgi:1983sy,Holdom:1985ag}, DM undergoes self-interaction mediated by a massive spin-1 gauge boson. This massive gauge boson, termed a ``dark photon''($A^\prime$), is similar to a SM photon except that it has a nonzero mass, $M_{A^\prime}$, and mediates a ``hidden'' $U(1)_D$ symmetry.  If a dark matter doublet exists which exhibits both SM hypercharge and DM hidden charge, the dark photon and the SM photon may undergo kinetic mixing, proving a bridge between the hidden sector and the Standard Model. Based on this concept, a number of potential schemes for detecting DM candidates and the dark photon have been proposed~\cite{Battaglieri:2017aum,Filippi:2020kii}.

\begin{figure*}[]
    \centering
    \includegraphics[width=0.9\linewidth]{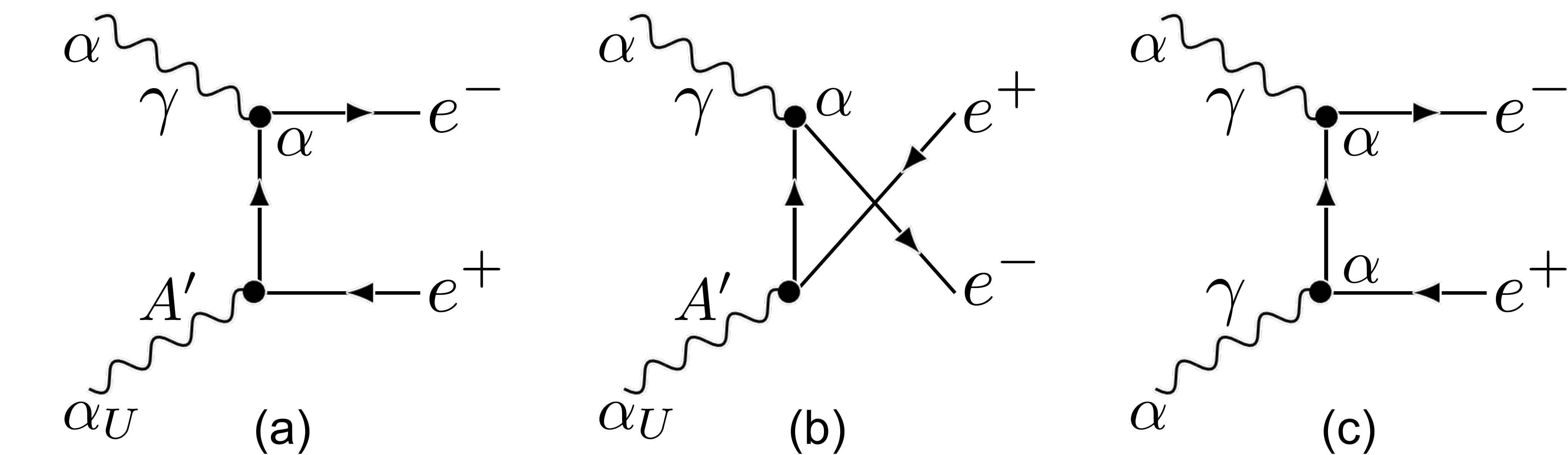}
    \caption{ Feynman diagrams illustrating photon and dark photon $(A^\prime)$ fusion processes in ultra-peripheral heavy ion collisions. The process involving one dark photon and one real photon, $\gamma A^\prime \rightarrow e^+e^-$, is shown in (a) and (b) where $\alpha_{_{U}}$ represents the dark photon coupling strength to charged standard model particles. The process involving standard model photons, $\gamma\gamma \rightarrow e^+e^-$, is shown in (c). The (dark) photons are manifest from the highly Lorentz-contracted electromagnetic field of the colliding nuclei. }
    \label{fig:process}
\end{figure*}

Much of what we know about the properties of DM comes from observations of the behavior of particles in the halo of the Milky-Way galaxy~\cite{Battaglieri:2017aum}.  Limits on the dark photon's mass and coupling to SM particles have also been set by accelerator experiments, which have the advantage of being independent of astrophysical uncertainties. The particles produced in accelerator experiments are also relativistic, which in the context of studying DM means that there are minimal contributions from velocity dependent mutual interactions between SM and DM particles~\cite{Filippi:2020kii}.   Some of these accelerator based experiments used more direct methods to set limits on the dark photon, including searching for resonances in dilepton pairs~\cite{BaBar:2014zli,NA482:2015wmo,Merkel:2014avp,PHENIX:2014duq,HADES:2013nab,HPS:2018xkw,LHCb:2019vmc}  or reinterpreting data from fixed target or neutrino experiments~\cite{Riordan:1987aw, Bjorken:1988as, Bross:1989mp}.
Indirect methods have also been used, such as searching for missing mass in exclusive collisions~\cite{BaBar:2017tiz, NA62:2019meo} or looking for missing energy or momentum in fixed target experiments~\cite{Banerjee:2019pds,LDMX:2018cma}. 

In this paper, we propose a new approach to search for and set limits on the dark photon parameters using the recent polarization measurement of the Breit-Wheeler process in ultra-peripheral collisions (UPC) from the STAR experiment at RHIC~\cite{STAR:2019wlg}. In 1934, Breit and Wheeler proposed that two real photons could interact to produce an electron-positron pair. However, this purely electromagnetic process went nearly a century without realization due to the smallness of the interaction cross section and the difficulty in obtaining sufficiently intense photon sources to overcome the energy threshold needed to produce an electron and a positron. However, heavy ions traveling at ultra-relativistic speeds produce one of the strongest electromagnetic fields in the known Universe~\cite{Bzdak:2011yy,Voronyuk:2011jd,Deng:2012pc,Roy:2015coa,Siddique:2021smf} and provide a viable source of photons for achieving the Breit-Wheeler process~\cite{Brandenburg:2022tna} and other light-by-light interactions~\cite{CMS:2018erd,ATLAS:2019azn,dEnterria:2013zqi}.  According to the Equivalent Photon Approximation (EPA), these highly Lorentz-contracted electromagnetic fields can be quantized as a source of photons~\cite{vonWeizsacker:1934nji, Williams:1934ad}.  The EPA has been found to hold for ultra-peripheral collisions at RHIC and LHC energies~\cite{Wang:2022ihj}.
 
Since these highly-Lorentz contracted electromagnetic fields are among the strongest ever observed, they reach high enough  energy densities that it is possible for them to manifest as dark photons with a mass of a few GeV or lower at RHIC and potentially higher at the LHC.
With this unique source of luminous and intense SM and DM photons, the time-reverse of the non-resonant production process can be achieved.  This is essentially the analog of the Breit-Wheeler process but with one or more SM photons replaced by a DM photon. In this case, a dark photon manifests from the field of one ion and may fuse with another (dark) photon manifested from other the ion's field, producing an observable $e^+e^-$ pair.  
Figs.~\ref{fig:process} (a) and (b) show Feynman diagrams for the case where one dark photon is manifested resulting in $\gamma A^\prime$ fusion  compared the Breit-Wheeler process in Figs.~\ref{fig:process} (c) where the $e^+e^-$ pair come from the fusion of two real photons. 
Since real photons are massless, their one unit of spin may only be carried in the $J_Z=\pm1$ states, while the $J_Z=0$ state is forbidden. The restriction on $J_Z$ for real photons compared to virtual photons, results in several characteristic and distinct signatures of the Breit-Wheeler process compared to other processes mediated by massive (virtual) photons. 

The STAR collaboration observed a significant $\cos{4\phi}$ azimuthal ($\phi$) asymmetry in the angle of the decay positron compared to the $e^+e^-$ pair momentum - an unmistakable characteristic of the Breit-Wheeler process. The observed $\cos{4\phi}$ modulation comes from the initial spin of the system being transferred into the final state $e^+e^-$ as orbital angular momentum.  This observation is related to vacuum birefringence~\cite{Heisenberg:1936nmg}, a phenomenon in which the QED vacuum behaves as a polarizable medium in the presence of a strong background electromagnetic field, leading to a splitting of the wave function of light as it passes through. The observation of this phenomenon offers a new avenue in the search for a massive dark photon since any mass in the initial colliding photons would allow contributions from the $J_Z=0$ state and directly affect the observed final-state azimuthal asymmetry. Most importantly, this technique is sensitive to the ratio of contributions from the various $J_Z$ states of the colliding photons, making it less susceptible to the often large experimental uncertainties on the total cross section measurement. 

\section{Dataset and event selection}
The STAR collaboration measured the Breit-Wheeler (BW) process in UPCs using $\rm{Au}+\rm{Au}$ collision data collected in 2010 from collisions with $\sqrt{s_{_{NN}}} = 200~\rm{GeV}$~\cite{STAR:2019wlg}. STAR detectors were used to trigger candidate BW events which contained an exclusive $e^+e^-$ pairs in conjunction with forward and backward activity in the Zero Degree Calorimeters (ZDC)~\cite{Adler:2000bd}. Activity in the ZDCs can result from Mutual Coulomb Excitation, in which one or both nuclei become excited by exchanging virtual photons before subsequently breaking apart~\cite{Bertulani:2005ru,Baltz:2009jk}. Beam-energy neutrons from this dissociative process, detected in the forward and backward ZDCs, provide a convenient signature for selecting exclusive mid-rapidity processes, such as the Breit-Wheeler process.

From the triggered events, $e^+e^-$ pairs were measured in the invariant mass range of $0.4<M_{ee}<2.6~\rm{GeV}$. The observation of a smooth and featureless invariant mass distribution, with no contribution from the production and subsequent decay of vector mesons, is expected for the BW process since two real photons cannot produce a single spin-$1$ meson~\cite{Brodsky:1981rp}. Besides the invariant mass distribution, the momentum distribution of the $e^+e^-$ pairs was also measured and compared with the expectations for the BW process.  The transverse momentum distribution of the $e^+e^-$ pairs was found to be consistent with QED calculations~\cite{Zha:2018tlq} displaying the characteristic peak in production cross section at $P_\perp \approx 20~\rm{MeV}$ expected for photons originating from the coherent field of a heavy ion.  Furthermore, the distribution of electron and positron momentum with respect to the polar angle displayed a preferential alignment with the beam direction, as expected for real photon collisions. 
This distribution was inconsistent with the production of $e^+e^-$ from virtual photons, which would be isotropic with respect to the polar angle.  

The fiducial cross section was measured for UPCs where one or more neutrons was emitted in both beam directions.  For exclusively produced $e^+e^-$ pairs with $0.4 < M_{ee} < 2.6~\rm{GeV}$ and $P_\perp < 0.1~\rm{GeV}$, the fiducial cross section was found to be $\sigma_{\rm{BW}} = 261 \pm 4$ (stat.) $\pm~13$ (syst.) $\pm~34$ (scale uncertainty) $\mu b$.  The total scale uncertainty is 13\% and is dominated by the uncertainty on the luminosity~\cite{STAR:2019wlg}.

As previously mentioned, these measurements also demonstrated an effect related to vacuum birefringence through the measurement of the difference in the azimuthal angle of the decay positron's momentum compared to the $e^+e^-$ momentum, as shown in Fig.~\ref{fig:azimuthal}.  In the Breit-Wheeler process, the two colliding spin-$1$ photons can only have a total angular momentum projection, $J_Z$, of $0$ or $\pm2$ since individual real photons are forbidden to inhabit the $J_Z=0$ state. When the two-photon system has $J_Z = 0$, the produced $e^+e^-$ pairs inherit no net orbital angular momentum and therefore display an isotropic azimuthal distribution. In the other case, when $J_Z = \pm2$, the produced pairs inherit two units of spin. Because the electron and positron are spin-$1/2$ particles and are predominately produced in the singlet state, 
the two units of spin are encoded into the momentum distribution of the $e^+e^-$ pairs as orbital angular momentum.  The measured azimuthal distribution was fit with the functional form 
\begin{equation}
 f(\phi) = C\big( 1 + A_{2\phi}\cos{2\phi} + A_{4\phi}\cos{4\phi} \big).
    \label{FitFunction}
\end{equation}
Here the $\cos{4\phi}$ term corresponds to the unique contribution from two real photon system being in the $J_Z = 2$ sate. The $|A_{4\phi}|$ coefficient was found to be large, $(16.8\pm 2.5)\%$, and consistent with a QED calculation that includes the effects of the photon polarization and the impact parameter dependence of the photon flux~\cite{Li:2019sin}. As expected for the BW process, the $|A_{2\phi}|$ coefficient was found to be consistent with zero, $ (2.0 \pm 2.4)\%$.  The uncertainty on this quantity  is dominated by limited statistics and also systematic uncertainties associated with reconstructing charged particle tracks at STAR~\cite{STAR:2019wlg}.  

A significant $\cos{2\phi}$ contribution would be expected if either of the two colliding photons could be in the $J_Z = 0$ state.  In the case of a dark photon interacting with a real photon to produce an $e^+e^-$ pair, the composite spin could be in the state of $J_Z=1$ and therefore a nonzero contribution to the $\cos{2\phi}$ distribution is possible. Therefore, we can utilize the relative amplitude between the measured $|A_{2\phi}|$ and $|A_{4\phi}|$ components of the azimuthal distribution to search for dark photon interactions, independent of the scale uncertainty of the total cross section.

\section{Searching for Dark Photon Interactions in Ultra-Peripheral Collisions}
The dark photon ($A^\prime$), a massive spin-1 gauge boson responsible for mediating the hidden sector $U(1)_D$, is expected to have similar properties as the SM photon except that it has a nonzero mass, $M_{A^\prime}$.  Furthermore, it may have a coupling constant to charged SM particles, $\alpha_{_{U}}=\epsilon^2\alpha$, which is expected to be small in comparison to the fine structure constant $\alpha$ ($\alpha_{_{U}} < 10^{-4} \times \alpha$), since the dark photon has yet to be observed~\cite{Holdom:1985ag}. An interaction consisting of a dark photon colliding with a real photon, $\gamma A^\prime\rightarrow e^+e^-$, is expected to differ from the Breit-Wheeler process based on these properties.

For unpolarized collisions, the only difference between $\gamma\gamma\rightarrow e^+e^-$ and $\gamma A^\prime\rightarrow e^+e^-$ at leading order is that one set of SM vertices has been replaced with dark photon vertices, as shown in Figs.~\ref{fig:process}(a) and (b). In the minimal version of the model~\cite{PHENIX:2014duq} and ignoring the branching ratio ($\sim100-50\%)$~\cite{HADES:2013nab}, the coupling of $A^\prime$ to the dielecton pair~\cite{Reece:2009un} is on the order of unity 
with possible off-shell mass. Although most of the previous and planned visual searches~\cite{Filippi:2020kii, LHCb:2019vmc} for such an object require it to be long-lived with decay, our search approach uses dark photon as the mediator~\cite{fayet2017light} with no requirement on its lifetime. The ratio of the cross section of this dark photon process, $\sigma_{A^\prime \gamma}$, to the cross section of Breit-Wheeler process, $\sigma_{\gamma\gamma}$, is proportional to $\alpha_{_{U}} / \alpha^2$.  Considering this dark photon process in addition to the SM QED cross section would result in an increase in the overall cross section for $e^+e^-$ pairs by some percentage.  Thus we use the total uncertainty of the measured fiducial cross section of the Breit-Wheeler process from Ref~\cite{STAR:2019wlg} to reject certain values of $\sigma_{A^\prime \gamma}$ with 90\% confidence, 

\begin{equation}
\frac{\sigma_{A^\prime \gamma}}{ \sigma_{\gamma\gamma}} 
   = \frac{\alpha_{_{U}}}{\alpha^2}= \frac{\epsilon^2}{\alpha}> 1.645 \times 13\%.
\label{crossSectionLimit}
\end{equation}

\begin{figure}[]
    \centering
    \includegraphics[width=1.0\linewidth]{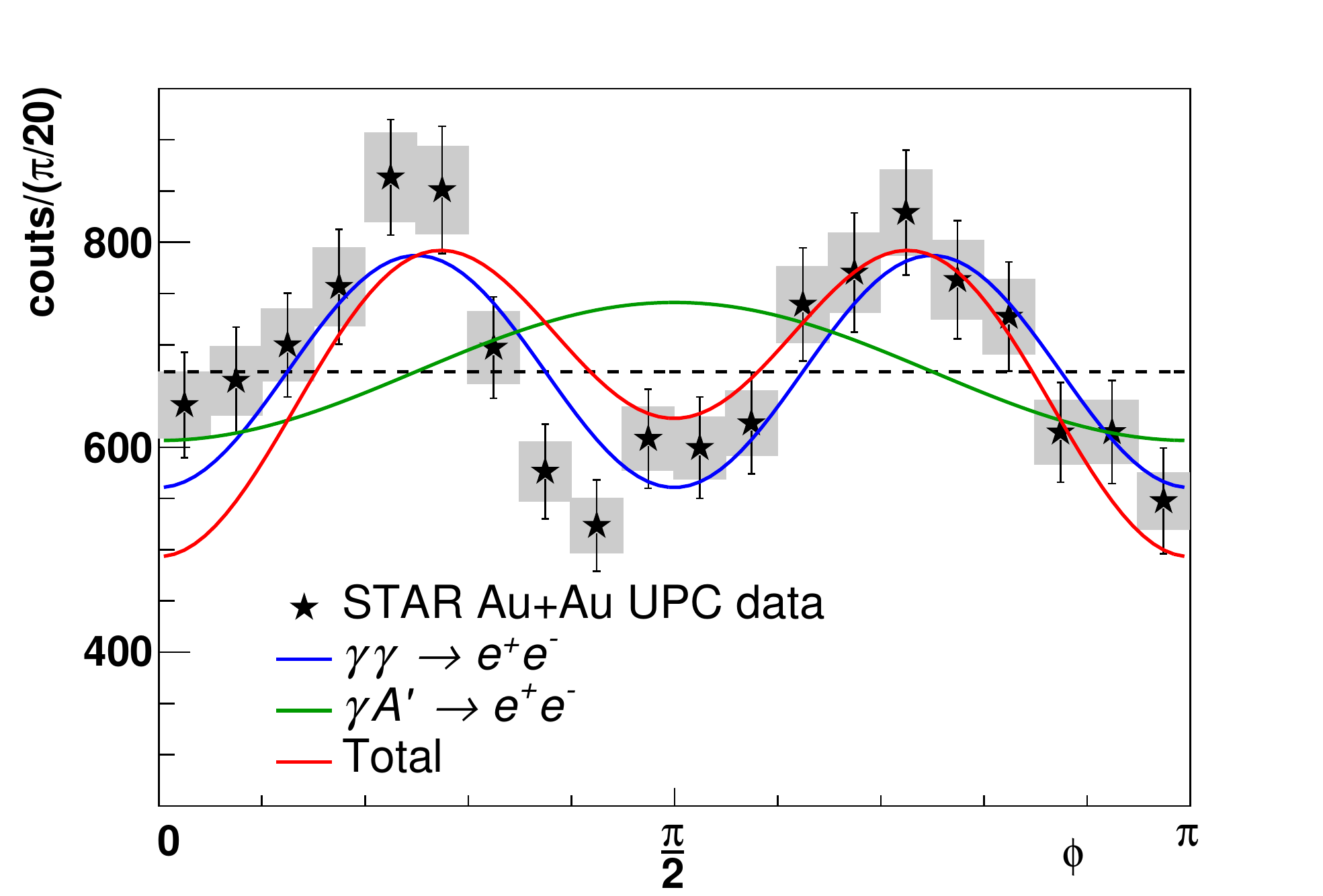}
    \caption{The azimuthal distribution for $e^+e^-$ pairs produced in ultra-peripheral $\rm{Au}+\rm{Au}$ collisions at $\sqrt{s_{_{NN}}} = 200~\rm{GeV}$. The data are reproduced from Ref~\cite{STAR:2019wlg} showing $e^+e^-$ pairs with $0.45 < M_{ee} < 0.76~\rm{GeV}$ and $P_\perp < 0.1~\rm{GeV}$. The blue curve shows the contribution from polarized $\gamma\gamma \rightarrow e^+e^-$ which because the photons are massless has no contribution from the $\cos{2\phi}$ term in Eq.~\ref{FitFunction}.  
    The green curve shows the $\cos{2\phi}$ contribution, with $|A_{2\phi}| = 10\%$, for a hypothetical dark photon signal with $M_{A^\prime} = 0.5$ GeV and $\epsilon^2=2.0\times 10^{-5}$. The red curve shows the sum of these contributions. }
    \label{fig:azimuthal}
\end{figure}

The non-zero mass of the dark photon allows it to be found in the $J_Z = 0$ state without breaking gauge invariance.  
This means that a dark photon scattering off of another photon can be in the composite $J_Z = 1$ state, which will lead to a non-zero amplitude for the $\cos{2\phi}$ term in Eq.~\ref{FitFunction} from the dominant cross terms between the mixing amplitudes. Fig.~\ref{fig:azimuthal} shows how this curve would look if this contribution were larger than allowed by the data, $|A_{2\phi}| = 10\%$. 

The more massive the dark photon, the more likely the system is to be in this state. Ref.~\cite{H1:2009cml} found that for vector meson production in diffractive electron-proton collisions, the ratio of the cross sections for when electric field of the virtual photon is longitudinally polarized versus transverse is $\sigma_L/\sigma_T = Q^2/M_v^2$, where $Q$ is the photon virtuality and $M_v$ is the mass of the vector meson. For $\gamma A^\prime\rightarrow e^+e^-$, this ratio would become $\sigma_L/\sigma_T = M_{A^{\prime}}^2/E^2$, were $E$ is the energy of the dark photon, $E=M_{ee}/2$. The likelihood of the $\cos{2\phi}$ term comes from the interference term between $\gamma A^\prime\rightarrow e^+e^-$ process (Fig.~\ref{fig:process}a) and  $\gamma\gamma\rightarrow e^+e^-$ (Fig.~\ref{fig:process}c)  with the relative strength determined by the size of the coupling constant $\sqrt{\alpha_{_{U}}}\alpha=\epsilon\alpha^{3/2}$ compared to $\alpha^2$~\cite{STAR:2019wlg,Li:2019sin}. The other term is the interference between the $\gamma A^\prime\rightarrow e^+e^-$ process (Fig.~\ref{fig:process}a) and  $\gamma A^\prime\rightarrow e^-e^+$ (Fig.~\ref{fig:process}b)  with the relative strength determined by the size of the coupling constant $\alpha_{_{U}}=\epsilon^2\alpha$ compared to $\alpha^2$. For small value of $\epsilon^2$ (much smaller than $\alpha$), the dominant sensitivity comes from the $\gamma A^\prime\rightarrow e^+e^-$ and  $\gamma\gamma\rightarrow e^+e^-$ interference:  $\epsilon^2/\alpha+\epsilon/\sqrt{\alpha}\simeq \epsilon/\sqrt{\alpha}$. Combining these properties together, we can set limits on the mass and mixing parameter by using the measured uncertainty on the $A_{2\phi}$ coefficient from Ref~\cite{STAR:2019wlg} via:

\begin{equation}
    A_{2\phi}(\epsilon, M_{A^\prime})     \simeq
    {\left(\frac{\epsilon^2}{\alpha}\right)^{1/2}}
    \left(2\frac{M_{A^\prime}}{M_{ee}}\right)^2
\label{spinExclusion1}
\end{equation}
To match with previous work, this study uses a confidence level of $90\%$ for the sensitivity limit.  In these processes, with $e^+e^-$ pairs produced near mid-rapidity we are able to probe dark photon masses of approximately $M_{A^\prime} \approx M_{ee} / 2.0$ since the pairs are produced with a negligible transverse momentum.
Therefore, we are only sensitive to dark photons with masses
\begin{equation}
    M_{A^\prime} \lesssim \frac{M_{ee} + \langle P_\perp \rangle}{2},
\label{spinExclusion2}
\end{equation}
where $\langle P_\perp \rangle \approx 40~\rm{MeV}$.

\begin{figure}[ht]
    \centering
    \includegraphics[width=1.0\linewidth]{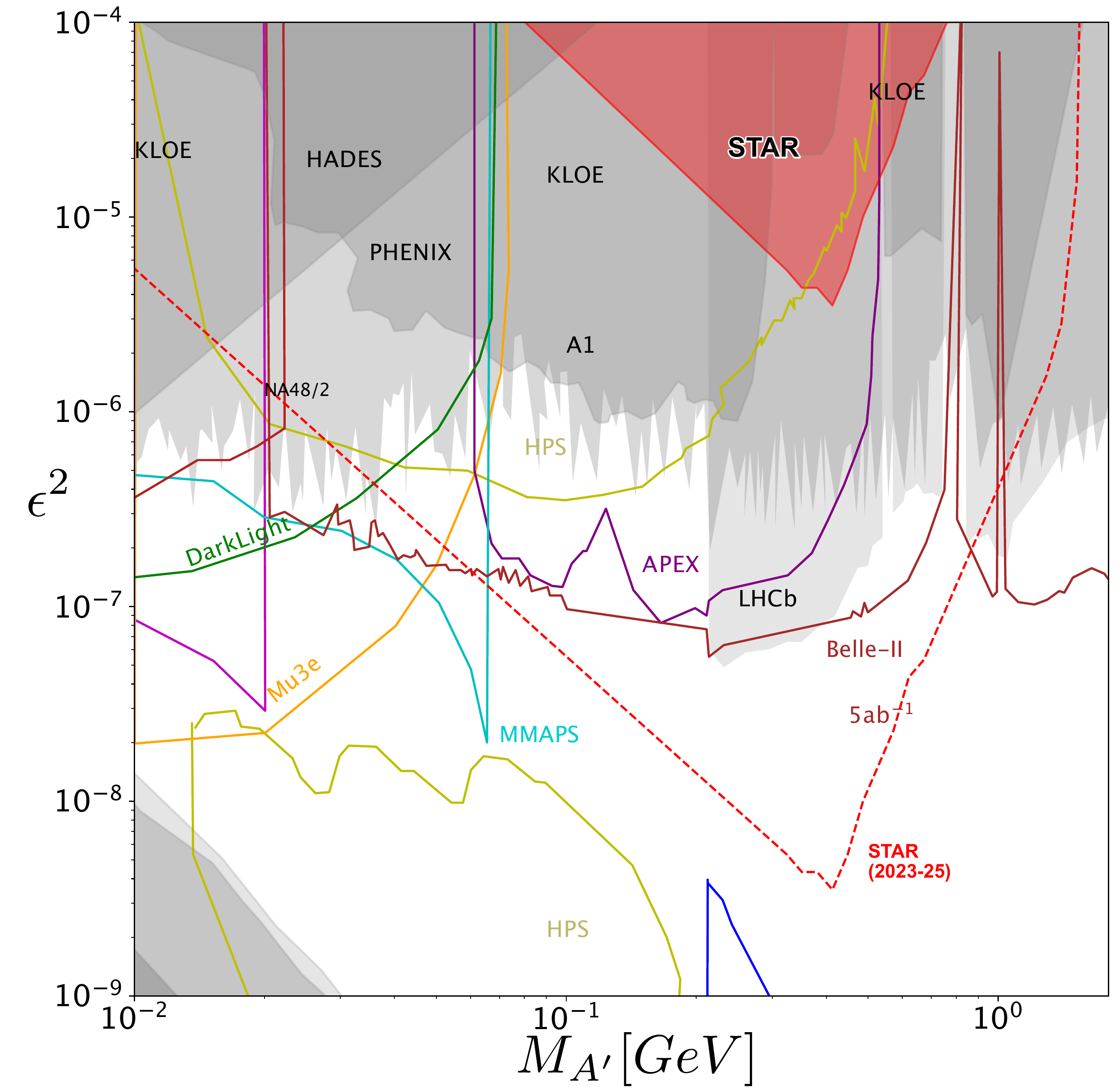}
    \caption{Exclusion plot for the existing bounds of the dark photon mass, $M_{A^\prime}$, and  mixing parameter to SM particles, $\epsilon^2$. The solid red region represents the region of phase space that is excluded by this result.  The dotted red line represents the estimated limit from future data.  The other curves come from Ref.~\cite{Filippi:2020kii,LHCb:2019vmc} where shaded gray areas correspond to limits set by past experiments and the colored outlines correspond to the predicted limits of ongoing or future experiments. These regions are excluded with 90\% confidence.}
    \label{fig:limits}
\end{figure}

Fig.~\ref{fig:limits} shows limits set on the dark photon mass ($M_{A^\prime}$) and SM mixing parameter ($\epsilon^2$) from this study, along with ruled-out parameter regions rejected by other experiments at the 90\% confidence.  
The gray shaded areas correspond to regions in phase space which have been excluded by other experiments, and the colored outlines correspond to predictions for limits to be set by future or ongoing experiments~\cite{Filippi:2020kii,LHCb:2019vmc}.  The red shaded area corresponds to the region excluded by this study using the Breit-Wheeler measurement by STAR from Ref.~\cite{STAR:2019wlg}.  
In the $0.1 \lessapprox M_{A^\prime} < 1~\rm{GeV}$ mass region, the excluded phase space takes on a triangular shape, which pushes our exclusion curve down to $\epsilon^2 < 3.5 \times 10^{-6}$ at its lowest value at $M_{A^\prime} \approx  0.45$ GeV. The limits are set by the angular distribution measurement in Eq.~\ref{spinExclusion1} and~\ref{spinExclusion2} which provides far more stringent constraints than the measured total cross section alone.

The limits in this study utilize the STAR Breit-Wheeler measurement performed with $23$ million UPC triggered events from $\rm{Au}+\rm{Au}$ collision at $\sqrt{s_{_{NN}}} = 200~\rm{GeV}$ collected in 2010~\cite{STAR:2019wlg}.
Future data taking at RHIC is expected to provide as much as $1000$ times more UPC triggered events collected in the years between 2023 to 2025~\cite{BUR}. For this reason, we present in Fig.~\ref{fig:limits} the predicted improvement in sensitivity expected from this procedure with $1000$ times more UPC data, which corresponds to the red dotted line. STAR also has large data sets from $\rm{Au}+\rm{Au}$ collisions at $\sqrt{s_{_{NN}}} = 200~\rm{GeV}$ taken in 2014 and 2016 that together contain a total of 2 trillion minimum bias collisions~\cite{BUR}. 
The $1000\times$ improvement in statistical precision pushes our exclusion at the $90\%$ confidence level down to $\epsilon^2 < 3.5 \times 10^{-6}$ at $M_{A^\prime} \approx  0.45$ GeV.
In the approach described in the current draft, we ignore the final-state Sudakov radiation~\cite{Klein:2020jom}. Once the sensitivity is improved to be better than $\epsilon^2<\alpha^3\simeq3\times10^{-7}$, a finite $\cos{2\phi}$ from the QED baseline due to these higher-order effects may be necessary. 
With these future data sets and the significant increase in statistics, the sensitivity of this approach should improve into regions which have not yet been fully explored by other experiments.

\section{Summary}
The STAR measurement of the  Breit-Wheeler process in UPCs from Ref.~\cite{STAR:2019wlg} was used to search for signatures of the dark photon through photon fusion 
in UPCs. 
In the absence of a positive signal, limits have been set on the dark photon mass ($M_{A^\prime}$) and the mixing parameter to SM particles ($\epsilon^2$).  
This was done through the measurement of the fiducial cross section and through the measured angular distribution of the produced $e^+e^-$ pairs.  
The non-zero mass of the dark photon, a hypothetical spin-1 gauge boson mediating the $U(1)_D$ hidden sector, allows it to inhabit the $J_Z=0$ state unlike real SM photons.  
For this reason, the $\gamma A^\prime \rightarrow e^+e^-$ process would result in a measurably distinct $\cos2\phi$ azimuthal asymmetry compared to the $\gamma\gamma\rightarrow e^+e^-$ involving real photons.  
This new technique in the search for dark photons allows us to search in continuous mass range without requirements on the Dark Photon lifetime and to place better constraints on $\epsilon$ than using the measured cross section alone.  It has the advantage of being independent of the often large experimental uncertainties on the luminosity and therefore the total measured cross section. 
The current limits are comparable to what has been measured at other experiments and fill in the gaps among the know particle masses where other searches are not sensitive to.  Using this technique more stringent limits on values of $\epsilon^2$ over the same range in $M_{A^\prime}$ will be possible with the additional data STAR has collected (2014, 2016) or plans to collect (2023, 2025).
We expect these limits to help inform future measurements of UPCs at STAR and to motivate further theoretical explorations of the dark photon.  
The nature, origin, and behavior of Dark Matter remain some of the most crucial open mysteries in modern physics, and the dark photon is a compelling mechanism with explanatory power that may provide an avenue for observation through Standard Model interactions.  

\section{acknowledgements} 

The authors thank Dr. Matheus Hostert (University of Minnesota) for comments on our draft leading to significant revision of the paper. We acknowledge many of the STAR collaborators for the fruitful discussions and the BNL Education and Outreach Department for the High School Summer Research Program. 
This work was funded by the U.S. DOE Office of Science under contract Nos. DE-SC0012704, DE-FG02-10ER41666, and DE-AC02-98CH10886 and by the National Natural Science Foundation of China 
under Grant Nos. 12075139. 

\bibliography{bibliography.bib}

\begin{thebibliography}{49}%
\makeatletter
\providecommand \@ifxundefined [1]{%
 \@ifx{#1\undefined}
}%
\providecommand \@ifnum [1]{%
 \ifnum #1\expandafter \@firstoftwo
 \else \expandafter \@secondoftwo
 \fi
}%
\providecommand \@ifx [1]{%
 \ifx #1\expandafter \@firstoftwo
 \else \expandafter \@secondoftwo
 \fi
}%
\providecommand \natexlab [1]{#1}%
\providecommand \enquote  [1]{``#1''}%
\providecommand \bibnamefont  [1]{#1}%
\providecommand \bibfnamefont [1]{#1}%
\providecommand \citenamefont [1]{#1}%
\providecommand \href@noop [0]{\@secondoftwo}%
\providecommand \href [0]{\begingroup \@sanitize@url \@href}%
\providecommand \@href[1]{\@@startlink{#1}\@@href}%
\providecommand \@@href[1]{\endgroup#1\@@endlink}%
\providecommand \@sanitize@url [0]{\catcode `\\12\catcode `\$12\catcode
  `\&12\catcode `\#12\catcode `\^12\catcode `\_12\catcode `\%12\relax}%
\providecommand \@@startlink[1]{}%
\providecommand \@@endlink[0]{}%
\providecommand \url  [0]{\begingroup\@sanitize@url \@url }%
\providecommand \@url [1]{\endgroup\@href {#1}{\urlprefix }}%
\providecommand \urlprefix  [0]{URL }%
\providecommand \Eprint [0]{\href }%
\providecommand \doibase [0]{http://dx.doi.org/}%
\providecommand \selectlanguage [0]{\@gobble}%
\providecommand \bibinfo  [0]{\@secondoftwo}%
\providecommand \bibfield  [0]{\@secondoftwo}%
\providecommand \translation [1]{[#1]}%
\providecommand \BibitemOpen [0]{}%
\providecommand \bibitemStop [0]{}%
\providecommand \bibitemNoStop [0]{.\EOS\space}%
\providecommand \EOS [0]{\spacefactor3000\relax}%
\providecommand \BibitemShut  [1]{\csname bibitem#1\endcsname}%
\let\auto@bib@innerbib\@empty
\bibitem [{\citenamefont {Zwicky}(1933)}]{Zwicky:1933gu}%
  \BibitemOpen
  \bibfield  {author} {\bibinfo {author} {\bibfnamefont {F.}~\bibnamefont
  {Zwicky}},\ }\bibfield  {title} {\enquote {\bibinfo {title} {{Die
  Rotverschiebung von extragalaktischen Nebeln}},}\ }\href {\doibase
  10.1007/s10714-008-0707-4} {\bibfield  {journal} {\bibinfo  {journal} {Helv.
  Phys. Acta}\ }\textbf {\bibinfo {volume} {6}},\ \bibinfo {pages} {110--127}
  (\bibinfo {year} {1933})}\BibitemShut {NoStop}%
\bibitem [{\citenamefont {Rubin}\ \emph {et~al.}(1980)\citenamefont {Rubin},
  \citenamefont {Thonnard},\ and\ \citenamefont {Ford}}]{Rubin:1980zd}%
  \BibitemOpen
  \bibfield  {author} {\bibinfo {author} {\bibfnamefont {V.~C.}\ \bibnamefont
  {Rubin}}, \bibinfo {author} {\bibfnamefont {N.}~\bibnamefont {Thonnard}}, \
  and\ \bibinfo {author} {\bibfnamefont {W.~K.}\ \bibnamefont {Ford},
  \bibfnamefont {Jr.}},\ }\bibfield  {title} {\enquote {\bibinfo {title}
  {{Rotational properties of 21 SC galaxies with a large range of luminosities
  and radii, from NGC 4605 (R = 4kpc) to UGC 2885 (R = 122 kpc)}},}\ }\href
  {\doibase 10.1086/158003} {\bibfield  {journal} {\bibinfo  {journal}
  {Astrophys. J.}\ }\textbf {\bibinfo {volume} {238}},\ \bibinfo {pages} {471}
  (\bibinfo {year} {1980})}\BibitemShut {NoStop}%
\bibitem [{\citenamefont {Clowe}\ \emph {et~al.}(2006)\citenamefont {Clowe},
  \citenamefont {Bradac}, \citenamefont {Gonzalez}, \citenamefont {Markevitch},
  \citenamefont {Randall}, \citenamefont {Jones},\ and\ \citenamefont
  {Zaritsky}}]{Clowe:2006eq}%
  \BibitemOpen
  \bibfield  {author} {\bibinfo {author} {\bibfnamefont {Douglas}\ \bibnamefont
  {Clowe}}, \bibinfo {author} {\bibfnamefont {Marusa}\ \bibnamefont {Bradac}},
  \bibinfo {author} {\bibfnamefont {Anthony~H.}\ \bibnamefont {Gonzalez}},
  \bibinfo {author} {\bibfnamefont {Maxim}\ \bibnamefont {Markevitch}},
  \bibinfo {author} {\bibfnamefont {Scott~W.}\ \bibnamefont {Randall}},
  \bibinfo {author} {\bibfnamefont {Christine}\ \bibnamefont {Jones}}, \ and\
  \bibinfo {author} {\bibfnamefont {Dennis}\ \bibnamefont {Zaritsky}},\
  }\bibfield  {title} {\enquote {\bibinfo {title} {{A direct empirical proof of
  the existence of dark matter}},}\ }\href {\doibase 10.1086/508162} {\bibfield
   {journal} {\bibinfo  {journal} {Astrophys. J. Lett.}\ }\textbf {\bibinfo
  {volume} {648}},\ \bibinfo {pages} {L109--L113} (\bibinfo {year} {2006})},\
  \Eprint {http://arxiv.org/abs/astro-ph/0608407} {arXiv:astro-ph/0608407}
  \BibitemShut {NoStop}%
\bibitem [{\citenamefont {Adam}\ \emph {et~al.}(2016)\citenamefont {Adam} \emph
  {et~al.}}]{Planck:2015mrs}%
  \BibitemOpen
  \bibfield  {author} {\bibinfo {author} {\bibfnamefont {R.}~\bibnamefont
  {Adam}} \emph {et~al.} (\bibinfo {collaboration} {Planck}),\ }\bibfield
  {title} {\enquote {\bibinfo {title} {{Planck 2015 results. I. Overview of
  products and scientific results}},}\ }\href {\doibase
  10.1051/0004-6361/201527101} {\bibfield  {journal} {\bibinfo  {journal}
  {Astron. Astrophys.}\ }\textbf {\bibinfo {volume} {594}},\ \bibinfo {pages}
  {A1} (\bibinfo {year} {2016})},\ \Eprint {http://arxiv.org/abs/1502.01582}
  {arXiv:1502.01582 [astro-ph.CO]} \BibitemShut {NoStop}%
\bibitem [{\citenamefont {Moore}(1994)}]{Moore:1994yx}%
  \BibitemOpen
  \bibfield  {author} {\bibinfo {author} {\bibfnamefont {B.}~\bibnamefont
  {Moore}},\ }\bibfield  {title} {\enquote {\bibinfo {title} {{Evidence against
  dissipationless dark matter from observations of galaxy haloes}},}\ }\href
  {\doibase 10.1038/370629a0} {\bibfield  {journal} {\bibinfo  {journal}
  {Nature}\ }\textbf {\bibinfo {volume} {370}},\ \bibinfo {pages} {629}
  (\bibinfo {year} {1994})}\BibitemShut {NoStop}%
\bibitem [{\citenamefont {Okun}(1982)}]{Okun:1982xi}%
  \BibitemOpen
  \bibfield  {author} {\bibinfo {author} {\bibfnamefont {L.~B.}\ \bibnamefont
  {Okun}},\ }\bibfield  {title} {\enquote {\bibinfo {title} {{LIMITS OF
  ELECTRODYNAMICS: PARAPHOTONS?}}}\ }\href@noop {} {\bibfield  {journal}
  {\bibinfo  {journal} {Sov. Phys. JETP}\ }\textbf {\bibinfo {volume} {56}},\
  \bibinfo {pages} {502} (\bibinfo {year} {1982})}\BibitemShut {NoStop}%
\bibitem [{\citenamefont {Georgi}\ \emph {et~al.}(1983)\citenamefont {Georgi},
  \citenamefont {Ginsparg},\ and\ \citenamefont {Glashow}}]{Georgi:1983sy}%
  \BibitemOpen
  \bibfield  {author} {\bibinfo {author} {\bibfnamefont {Howard}\ \bibnamefont
  {Georgi}}, \bibinfo {author} {\bibfnamefont {Paul~H.}\ \bibnamefont
  {Ginsparg}}, \ and\ \bibinfo {author} {\bibfnamefont {S.~L.}\ \bibnamefont
  {Glashow}},\ }\bibfield  {title} {\enquote {\bibinfo {title} {{Photon
  Oscillations and the Cosmic Background Radiation}},}\ }\href {\doibase
  10.1038/306765a0} {\bibfield  {journal} {\bibinfo  {journal} {Nature}\
  }\textbf {\bibinfo {volume} {306}},\ \bibinfo {pages} {765--766} (\bibinfo
  {year} {1983})}\BibitemShut {NoStop}%
\bibitem [{\citenamefont {Holdom}(1986)}]{Holdom:1985ag}%
  \BibitemOpen
  \bibfield  {author} {\bibinfo {author} {\bibfnamefont {Bob}\ \bibnamefont
  {Holdom}},\ }\bibfield  {title} {\enquote {\bibinfo {title} {{Two U(1)'s and
  Epsilon Charge Shifts}},}\ }\href {\doibase 10.1016/0370-2693(86)91377-8}
  {\bibfield  {journal} {\bibinfo  {journal} {Phys. Lett. B}\ }\textbf
  {\bibinfo {volume} {166}},\ \bibinfo {pages} {196--198} (\bibinfo {year}
  {1986})}\BibitemShut {NoStop}%
\bibitem [{\citenamefont {Battaglieri}\ \emph {et~al.}(2017)\citenamefont
  {Battaglieri} \emph {et~al.}}]{Battaglieri:2017aum}%
  \BibitemOpen
  \bibfield  {author} {\bibinfo {author} {\bibfnamefont {Marco}\ \bibnamefont
  {Battaglieri}} \emph {et~al.},\ }\bibfield  {title} {\enquote {\bibinfo
  {title} {{US Cosmic Visions: New Ideas in Dark Matter 2017: Community
  Report}},}\ }in\ \href@noop {} {\emph {\bibinfo {booktitle} {{U.S. Cosmic
  Visions: New Ideas in Dark Matter}}}}\ (\bibinfo {year} {2017})\ \Eprint
  {http://arxiv.org/abs/1707.04591} {arXiv:1707.04591 [hep-ph]} \BibitemShut
  {NoStop}%
\bibitem [{\citenamefont {Filippi}\ and\ \citenamefont
  {De~Napoli}(2020)}]{Filippi:2020kii}%
  \BibitemOpen
  \bibfield  {author} {\bibinfo {author} {\bibfnamefont {Alessandra}\
  \bibnamefont {Filippi}}\ and\ \bibinfo {author} {\bibfnamefont {Marzio}\
  \bibnamefont {De~Napoli}},\ }\bibfield  {title} {\enquote {\bibinfo {title}
  {{Searching in the dark: the hunt for the dark photon}},}\ }\href {\doibase
  10.1016/j.revip.2020.100042} {\bibfield  {journal} {\bibinfo  {journal} {Rev.
  Phys.}\ }\textbf {\bibinfo {volume} {5}},\ \bibinfo {pages} {100042}
  (\bibinfo {year} {2020})},\ \Eprint {http://arxiv.org/abs/2006.04640}
  {arXiv:2006.04640 [hep-ph]} \BibitemShut {NoStop}%
\bibitem [{\citenamefont {Lees}\ \emph {et~al.}(2014)\citenamefont {Lees} \emph
  {et~al.}}]{BaBar:2014zli}%
  \BibitemOpen
  \bibfield  {author} {\bibinfo {author} {\bibfnamefont {J.~P.}\ \bibnamefont
  {Lees}} \emph {et~al.} (\bibinfo {collaboration} {BaBar}),\ }\bibfield
  {title} {\enquote {\bibinfo {title} {{Search for a Dark Photon in $e^+e^-$
  Collisions at BaBar}},}\ }\href {\doibase 10.1103/PhysRevLett.113.201801}
  {\bibfield  {journal} {\bibinfo  {journal} {Phys. Rev. Lett.}\ }\textbf
  {\bibinfo {volume} {113}},\ \bibinfo {pages} {201801} (\bibinfo {year}
  {2014})},\ \Eprint {http://arxiv.org/abs/1406.2980} {arXiv:1406.2980
  [hep-ex]} \BibitemShut {NoStop}%
\bibitem [{\citenamefont {Batley}\ \emph {et~al.}(2015)\citenamefont {Batley}
  \emph {et~al.}}]{NA482:2015wmo}%
  \BibitemOpen
  \bibfield  {author} {\bibinfo {author} {\bibfnamefont {J.~R.}\ \bibnamefont
  {Batley}} \emph {et~al.} (\bibinfo {collaboration} {NA48/2}),\ }\bibfield
  {title} {\enquote {\bibinfo {title} {{Search for the dark photon in $\pi^0$
  decays}},}\ }\href {\doibase 10.1016/j.physletb.2015.04.068} {\bibfield
  {journal} {\bibinfo  {journal} {Phys. Lett. B}\ }\textbf {\bibinfo {volume}
  {746}},\ \bibinfo {pages} {178--185} (\bibinfo {year} {2015})},\ \Eprint
  {http://arxiv.org/abs/1504.00607} {arXiv:1504.00607 [hep-ex]} \BibitemShut
  {NoStop}%
\bibitem [{\citenamefont {Merkel}\ \emph {et~al.}(2014)\citenamefont {Merkel}
  \emph {et~al.}}]{Merkel:2014avp}%
  \BibitemOpen
  \bibfield  {author} {\bibinfo {author} {\bibfnamefont {H.}~\bibnamefont
  {Merkel}} \emph {et~al.},\ }\bibfield  {title} {\enquote {\bibinfo {title}
  {{Search at the Mainz Microtron for Light Massive Gauge Bosons Relevant for
  the Muon g-2 Anomaly}},}\ }\href {\doibase 10.1103/PhysRevLett.112.221802}
  {\bibfield  {journal} {\bibinfo  {journal} {Phys. Rev. Lett.}\ }\textbf
  {\bibinfo {volume} {112}},\ \bibinfo {pages} {221802} (\bibinfo {year}
  {2014})},\ \Eprint {http://arxiv.org/abs/1404.5502} {arXiv:1404.5502
  [hep-ex]} \BibitemShut {NoStop}%
\bibitem [{\citenamefont {Adare}\ \emph {et~al.}(2015)\citenamefont {Adare}
  \emph {et~al.}}]{PHENIX:2014duq}%
  \BibitemOpen
  \bibfield  {author} {\bibinfo {author} {\bibfnamefont {A.}~\bibnamefont
  {Adare}} \emph {et~al.} (\bibinfo {collaboration} {PHENIX}),\ }\bibfield
  {title} {\enquote {\bibinfo {title} {{Search for dark photons from neutral
  meson decays in $p + p$ and $d$ + Au collisions at $\sqrt{s_{NN}} =$ 200
  GeV}},}\ }\href {\doibase 10.1103/PhysRevC.91.031901} {\bibfield  {journal}
  {\bibinfo  {journal} {Phys. Rev. C}\ }\textbf {\bibinfo {volume} {91}},\
  \bibinfo {pages} {031901} (\bibinfo {year} {2015})},\ \Eprint
  {http://arxiv.org/abs/1409.0851} {arXiv:1409.0851 [nucl-ex]} \BibitemShut
  {NoStop}%
\bibitem [{\citenamefont {Agakishiev}\ \emph {et~al.}(2014)\citenamefont
  {Agakishiev} \emph {et~al.}}]{HADES:2013nab}%
  \BibitemOpen
  \bibfield  {author} {\bibinfo {author} {\bibfnamefont {G.}~\bibnamefont
  {Agakishiev}} \emph {et~al.} (\bibinfo {collaboration} {HADES}),\ }\bibfield
  {title} {\enquote {\bibinfo {title} {{Searching a Dark Photon with HADES}},}\
  }\href {\doibase 10.1016/j.physletb.2014.02.035} {\bibfield  {journal}
  {\bibinfo  {journal} {Phys. Lett. B}\ }\textbf {\bibinfo {volume} {731}},\
  \bibinfo {pages} {265--271} (\bibinfo {year} {2014})},\ \Eprint
  {http://arxiv.org/abs/1311.0216} {arXiv:1311.0216 [hep-ex]} \BibitemShut
  {NoStop}%
\bibitem [{\citenamefont {Adrian}\ \emph {et~al.}(2018)\citenamefont {Adrian}
  \emph {et~al.}}]{HPS:2018xkw}%
  \BibitemOpen
  \bibfield  {author} {\bibinfo {author} {\bibfnamefont {P.~H.}\ \bibnamefont
  {Adrian}} \emph {et~al.} (\bibinfo {collaboration} {HPS}),\ }\bibfield
  {title} {\enquote {\bibinfo {title} {{Search for a dark photon in
  electroproduced $e^{+}e^{-}$ pairs with the Heavy Photon Search experiment at
  JLab}},}\ }\href {\doibase 10.1103/PhysRevD.98.091101} {\bibfield  {journal}
  {\bibinfo  {journal} {Phys. Rev. D}\ }\textbf {\bibinfo {volume} {98}},\
  \bibinfo {pages} {091101} (\bibinfo {year} {2018})},\ \Eprint
  {http://arxiv.org/abs/1807.11530} {arXiv:1807.11530 [hep-ex]} \BibitemShut
  {NoStop}%
\bibitem [{\citenamefont {Aaij}\ \emph {et~al.}(2020)\citenamefont {Aaij} \emph
  {et~al.}}]{LHCb:2019vmc}%
  \BibitemOpen
  \bibfield  {author} {\bibinfo {author} {\bibfnamefont {Roel}\ \bibnamefont
  {Aaij}} \emph {et~al.} (\bibinfo {collaboration} {LHCb}),\ }\bibfield
  {title} {\enquote {\bibinfo {title} {{Search for $A'\to\mu^+\mu^-$
  Decays}},}\ }\href {\doibase 10.1103/PhysRevLett.124.041801} {\bibfield
  {journal} {\bibinfo  {journal} {Phys. Rev. Lett.}\ }\textbf {\bibinfo
  {volume} {124}},\ \bibinfo {pages} {041801} (\bibinfo {year} {2020})},\
  \Eprint {http://arxiv.org/abs/1910.06926} {arXiv:1910.06926 [hep-ex]}
  \BibitemShut {NoStop}%
\bibitem [{\citenamefont {Riordan}\ \emph {et~al.}(1987)\citenamefont {Riordan}
  \emph {et~al.}}]{Riordan:1987aw}%
  \BibitemOpen
  \bibfield  {author} {\bibinfo {author} {\bibfnamefont {E.~M.}\ \bibnamefont
  {Riordan}} \emph {et~al.},\ }\bibfield  {title} {\enquote {\bibinfo {title}
  {{A Search for Short Lived Axions in an Electron Beam Dump Experiment}},}\
  }\href {\doibase 10.1103/PhysRevLett.59.755} {\bibfield  {journal} {\bibinfo
  {journal} {Phys. Rev. Lett.}\ }\textbf {\bibinfo {volume} {59}},\ \bibinfo
  {pages} {755} (\bibinfo {year} {1987})}\BibitemShut {NoStop}%
\bibitem [{\citenamefont {Bjorken}\ \emph {et~al.}(1988)\citenamefont
  {Bjorken}, \citenamefont {Ecklund}, \citenamefont {Nelson}, \citenamefont
  {Abashian}, \citenamefont {Church}, \citenamefont {Lu}, \citenamefont {Mo},
  \citenamefont {Nunamaker},\ and\ \citenamefont {Rassmann}}]{Bjorken:1988as}%
  \BibitemOpen
  \bibfield  {author} {\bibinfo {author} {\bibfnamefont {J.~D.}\ \bibnamefont
  {Bjorken}}, \bibinfo {author} {\bibfnamefont {S.}~\bibnamefont {Ecklund}},
  \bibinfo {author} {\bibfnamefont {W.~R.}\ \bibnamefont {Nelson}}, \bibinfo
  {author} {\bibfnamefont {A.}~\bibnamefont {Abashian}}, \bibinfo {author}
  {\bibfnamefont {C.}~\bibnamefont {Church}}, \bibinfo {author} {\bibfnamefont
  {B.}~\bibnamefont {Lu}}, \bibinfo {author} {\bibfnamefont {L.~W.}\
  \bibnamefont {Mo}}, \bibinfo {author} {\bibfnamefont {T.~A.}\ \bibnamefont
  {Nunamaker}}, \ and\ \bibinfo {author} {\bibfnamefont {P.}~\bibnamefont
  {Rassmann}},\ }\bibfield  {title} {\enquote {\bibinfo {title} {{Search for
  Neutral Metastable Penetrating Particles Produced in the SLAC Beam Dump}},}\
  }\href {\doibase 10.1103/PhysRevD.38.3375} {\bibfield  {journal} {\bibinfo
  {journal} {Phys. Rev. D}\ }\textbf {\bibinfo {volume} {38}},\ \bibinfo
  {pages} {3375} (\bibinfo {year} {1988})}\BibitemShut {NoStop}%
\bibitem [{\citenamefont {Bross}\ \emph {et~al.}(1991)\citenamefont {Bross},
  \citenamefont {Crisler}, \citenamefont {Pordes}, \citenamefont {Volk},
  \citenamefont {Errede},\ and\ \citenamefont {Wrbanek}}]{Bross:1989mp}%
  \BibitemOpen
  \bibfield  {author} {\bibinfo {author} {\bibfnamefont {A.}~\bibnamefont
  {Bross}}, \bibinfo {author} {\bibfnamefont {M.}~\bibnamefont {Crisler}},
  \bibinfo {author} {\bibfnamefont {Stephen~H.}\ \bibnamefont {Pordes}},
  \bibinfo {author} {\bibfnamefont {J.}~\bibnamefont {Volk}}, \bibinfo {author}
  {\bibfnamefont {S.}~\bibnamefont {Errede}}, \ and\ \bibinfo {author}
  {\bibfnamefont {J.}~\bibnamefont {Wrbanek}},\ }\bibfield  {title} {\enquote
  {\bibinfo {title} {{A Search for Shortlived Particles Produced in an Electron
  Beam Dump}},}\ }\href {\doibase 10.1103/PhysRevLett.67.2942} {\bibfield
  {journal} {\bibinfo  {journal} {Phys. Rev. Lett.}\ }\textbf {\bibinfo
  {volume} {67}},\ \bibinfo {pages} {2942--2945} (\bibinfo {year}
  {1991})}\BibitemShut {NoStop}%
\bibitem [{\citenamefont {Lees}\ \emph {et~al.}(2017)\citenamefont {Lees} \emph
  {et~al.}}]{BaBar:2017tiz}%
  \BibitemOpen
  \bibfield  {author} {\bibinfo {author} {\bibfnamefont {J.~P.}\ \bibnamefont
  {Lees}} \emph {et~al.} (\bibinfo {collaboration} {BaBar}),\ }\bibfield
  {title} {\enquote {\bibinfo {title} {{Search for Invisible Decays of a Dark
  Photon Produced in ${e}^{+}{e}^{-}$ Collisions at BaBar}},}\ }\href {\doibase
  10.1103/PhysRevLett.119.131804} {\bibfield  {journal} {\bibinfo  {journal}
  {Phys. Rev. Lett.}\ }\textbf {\bibinfo {volume} {119}},\ \bibinfo {pages}
  {131804} (\bibinfo {year} {2017})},\ \Eprint
  {http://arxiv.org/abs/1702.03327} {arXiv:1702.03327 [hep-ex]} \BibitemShut
  {NoStop}%
\bibitem [{\citenamefont {Cortina~Gil}\ \emph {et~al.}(2019)\citenamefont
  {Cortina~Gil} \emph {et~al.}}]{NA62:2019meo}%
  \BibitemOpen
  \bibfield  {author} {\bibinfo {author} {\bibfnamefont {Eduardo}\ \bibnamefont
  {Cortina~Gil}} \emph {et~al.} (\bibinfo {collaboration} {NA62}),\ }\bibfield
  {title} {\enquote {\bibinfo {title} {{Search for production of an invisible
  dark photon in $\pi^0$ decays}},}\ }\href {\doibase 10.1007/JHEP05(2019)182}
  {\bibfield  {journal} {\bibinfo  {journal} {JHEP}\ }\textbf {\bibinfo
  {volume} {05}},\ \bibinfo {pages} {182} (\bibinfo {year} {2019})},\ \Eprint
  {http://arxiv.org/abs/1903.08767} {arXiv:1903.08767 [hep-ex]} \BibitemShut
  {NoStop}%
\bibitem [{\citenamefont {Banerjee}\ \emph {et~al.}(2019)\citenamefont
  {Banerjee} \emph {et~al.}}]{Banerjee:2019pds}%
  \BibitemOpen
  \bibfield  {author} {\bibinfo {author} {\bibfnamefont {D.}~\bibnamefont
  {Banerjee}} \emph {et~al.},\ }\bibfield  {title} {\enquote {\bibinfo {title}
  {{Dark matter search in missing energy events with NA64}},}\ }\href {\doibase
  10.1103/PhysRevLett.123.121801} {\bibfield  {journal} {\bibinfo  {journal}
  {Phys. Rev. Lett.}\ }\textbf {\bibinfo {volume} {123}},\ \bibinfo {pages}
  {121801} (\bibinfo {year} {2019})},\ \Eprint
  {http://arxiv.org/abs/1906.00176} {arXiv:1906.00176 [hep-ex]} \BibitemShut
  {NoStop}%
\bibitem [{\citenamefont {\r{A}kesson}\ \emph {et~al.}(2018)\citenamefont
  {\r{A}kesson} \emph {et~al.}}]{LDMX:2018cma}%
  \BibitemOpen
  \bibfield  {author} {\bibinfo {author} {\bibfnamefont {Torsten}\ \bibnamefont
  {\r{A}kesson}} \emph {et~al.} (\bibinfo {collaboration} {LDMX}),\ }\bibfield
  {title} {\enquote {\bibinfo {title} {{Light Dark Matter eXperiment
  (LDMX)}},}\ }\href@noop {} {\  (\bibinfo {year} {2018})},\ \Eprint
  {http://arxiv.org/abs/1808.05219} {arXiv:1808.05219 [hep-ex]} \BibitemShut
  {NoStop}%
\bibitem [{\citenamefont {Adam}\ \emph {et~al.}(2021)\citenamefont {Adam} \emph
  {et~al.}}]{STAR:2019wlg}%
  \BibitemOpen
  \bibfield  {author} {\bibinfo {author} {\bibfnamefont {Jaroslav}\
  \bibnamefont {Adam}} \emph {et~al.} (\bibinfo {collaboration} {STAR}),\
  }\bibfield  {title} {\enquote {\bibinfo {title} {{Measurement of $e^+e^-$
  Momentum and Angular Distributions from Linearly Polarized Photon
  Collisions}},}\ }\href {\doibase 10.1103/PhysRevLett.127.052302} {\bibfield
  {journal} {\bibinfo  {journal} {Phys. Rev. Lett.}\ }\textbf {\bibinfo
  {volume} {127}},\ \bibinfo {pages} {052302} (\bibinfo {year} {2021})},\
  \Eprint {http://arxiv.org/abs/1910.12400} {arXiv:1910.12400 [nucl-ex]}
  \BibitemShut {NoStop}%
\bibitem [{\citenamefont {Bzdak}\ and\ \citenamefont
  {Skokov}(2012)}]{Bzdak:2011yy}%
  \BibitemOpen
  \bibfield  {author} {\bibinfo {author} {\bibfnamefont {Adam}\ \bibnamefont
  {Bzdak}}\ and\ \bibinfo {author} {\bibfnamefont {Vladimir}\ \bibnamefont
  {Skokov}},\ }\bibfield  {title} {\enquote {\bibinfo {title} {{Event-by-event
  fluctuations of magnetic and electric fields in heavy ion collisions}},}\
  }\href {\doibase 10.1016/j.physletb.2012.02.065} {\bibfield  {journal}
  {\bibinfo  {journal} {Phys. Lett. B}\ }\textbf {\bibinfo {volume} {710}},\
  \bibinfo {pages} {171--174} (\bibinfo {year} {2012})},\ \Eprint
  {http://arxiv.org/abs/1111.1949} {arXiv:1111.1949 [hep-ph]} \BibitemShut
  {NoStop}%
\bibitem [{\citenamefont {Voronyuk}\ \emph {et~al.}(2011)\citenamefont
  {Voronyuk}, \citenamefont {Toneev}, \citenamefont {Cassing}, \citenamefont
  {Bratkovskaya}, \citenamefont {Konchakovski},\ and\ \citenamefont
  {Voloshin}}]{Voronyuk:2011jd}%
  \BibitemOpen
  \bibfield  {author} {\bibinfo {author} {\bibfnamefont {V.}~\bibnamefont
  {Voronyuk}}, \bibinfo {author} {\bibfnamefont {V.~D.}\ \bibnamefont
  {Toneev}}, \bibinfo {author} {\bibfnamefont {W.}~\bibnamefont {Cassing}},
  \bibinfo {author} {\bibfnamefont {E.~L.}\ \bibnamefont {Bratkovskaya}},
  \bibinfo {author} {\bibfnamefont {V.~P.}\ \bibnamefont {Konchakovski}}, \
  and\ \bibinfo {author} {\bibfnamefont {S.~A.}\ \bibnamefont {Voloshin}},\
  }\bibfield  {title} {\enquote {\bibinfo {title} {{(Electro-)Magnetic field
  evolution in relativistic heavy-ion collisions}},}\ }\href {\doibase
  10.1103/PhysRevC.83.054911} {\bibfield  {journal} {\bibinfo  {journal} {Phys.
  Rev. C}\ }\textbf {\bibinfo {volume} {83}},\ \bibinfo {pages} {054911}
  (\bibinfo {year} {2011})},\ \Eprint {http://arxiv.org/abs/1103.4239}
  {arXiv:1103.4239 [nucl-th]} \BibitemShut {NoStop}%
\bibitem [{\citenamefont {Deng}\ and\ \citenamefont
  {Huang}(2012)}]{Deng:2012pc}%
  \BibitemOpen
  \bibfield  {author} {\bibinfo {author} {\bibfnamefont {Wei-Tian}\
  \bibnamefont {Deng}}\ and\ \bibinfo {author} {\bibfnamefont {Xu-Guang}\
  \bibnamefont {Huang}},\ }\bibfield  {title} {\enquote {\bibinfo {title}
  {{Event-by-event generation of electromagnetic fields in heavy-ion
  collisions}},}\ }\href {\doibase 10.1103/PhysRevC.85.044907} {\bibfield
  {journal} {\bibinfo  {journal} {Phys. Rev. C}\ }\textbf {\bibinfo {volume}
  {85}},\ \bibinfo {pages} {044907} (\bibinfo {year} {2012})},\ \Eprint
  {http://arxiv.org/abs/1201.5108} {arXiv:1201.5108 [nucl-th]} \BibitemShut
  {NoStop}%
\bibitem [{\citenamefont {Roy}\ and\ \citenamefont {Pu}(2015)}]{Roy:2015coa}%
  \BibitemOpen
  \bibfield  {author} {\bibinfo {author} {\bibfnamefont {Victor}\ \bibnamefont
  {Roy}}\ and\ \bibinfo {author} {\bibfnamefont {Shi}\ \bibnamefont {Pu}},\
  }\bibfield  {title} {\enquote {\bibinfo {title} {{Event-by-event distribution
  of magnetic field energy over initial fluid energy density in $\sqrt{s_{\rm
  NN}}$= 200 GeV Au-Au collisions}},}\ }\href {\doibase
  10.1103/PhysRevC.92.064902} {\bibfield  {journal} {\bibinfo  {journal} {Phys.
  Rev. C}\ }\textbf {\bibinfo {volume} {92}},\ \bibinfo {pages} {064902}
  (\bibinfo {year} {2015})},\ \Eprint {http://arxiv.org/abs/1508.03761}
  {arXiv:1508.03761 [nucl-th]} \BibitemShut {NoStop}%
\bibitem [{\citenamefont {Siddique}\ \emph {et~al.}(2021)\citenamefont
  {Siddique}, \citenamefont {Sheng},\ and\ \citenamefont
  {Wang}}]{Siddique:2021smf}%
  \BibitemOpen
  \bibfield  {author} {\bibinfo {author} {\bibfnamefont {Irfan}\ \bibnamefont
  {Siddique}}, \bibinfo {author} {\bibfnamefont {Xin-Li}\ \bibnamefont
  {Sheng}}, \ and\ \bibinfo {author} {\bibfnamefont {Qun}\ \bibnamefont
  {Wang}},\ }\bibfield  {title} {\enquote {\bibinfo {title} {{Space-average
  electromagnetic fields and electromagnetic anomaly weighted by energy density
  in heavy-ion collisions}},}\ }\href {\doibase 10.1103/PhysRevC.104.034907}
  {\bibfield  {journal} {\bibinfo  {journal} {Phys. Rev. C}\ }\textbf {\bibinfo
  {volume} {104}},\ \bibinfo {pages} {034907} (\bibinfo {year} {2021})},\
  \Eprint {http://arxiv.org/abs/2106.00478} {arXiv:2106.00478 [nucl-th]}
  \BibitemShut {NoStop}%
\bibitem [{\citenamefont {Brandenburg}\ \emph {et~al.}(2022)\citenamefont
  {Brandenburg}, \citenamefont {Seger}, \citenamefont {Xu},\ and\ \citenamefont
  {Zha}}]{Brandenburg:2022tna}%
  \BibitemOpen
  \bibfield  {author} {\bibinfo {author} {\bibfnamefont {James~Daniel}\
  \bibnamefont {Brandenburg}}, \bibinfo {author} {\bibfnamefont {Janet}\
  \bibnamefont {Seger}}, \bibinfo {author} {\bibfnamefont {Zhangbu}\
  \bibnamefont {Xu}}, \ and\ \bibinfo {author} {\bibfnamefont {Wangmei}\
  \bibnamefont {Zha}},\ }\bibfield  {title} {\enquote {\bibinfo {title}
  {{Report on Progress in Physics: Observation of the Breit-Wheeler Process and
  Vacuum Birefringence in Heavy-Ion Collisions}},}\ }\href@noop {} {\
  (\bibinfo {year} {2022})},\ \Eprint {http://arxiv.org/abs/2208.14943}
  {arXiv:2208.14943 [hep-ph]} \BibitemShut {NoStop}%
\bibitem [{\citenamefont {Sirunyan}\ \emph {et~al.}(2019)\citenamefont
  {Sirunyan} \emph {et~al.}}]{CMS:2018erd}%
  \BibitemOpen
  \bibfield  {author} {\bibinfo {author} {\bibfnamefont {Albert~M}\
  \bibnamefont {Sirunyan}} \emph {et~al.} (\bibinfo {collaboration} {CMS}),\
  }\bibfield  {title} {\enquote {\bibinfo {title} {{Evidence for light-by-light
  scattering and searches for axion-like particles in ultraperipheral PbPb
  collisions at $\sqrt{s_\mathrm{NN}} =$ 5.02 TeV}},}\ }\href {\doibase
  10.1016/j.physletb.2019.134826} {\bibfield  {journal} {\bibinfo  {journal}
  {Phys. Lett. B}\ }\textbf {\bibinfo {volume} {797}},\ \bibinfo {pages}
  {134826} (\bibinfo {year} {2019})},\ \Eprint
  {http://arxiv.org/abs/1810.04602} {arXiv:1810.04602 [hep-ex]} \BibitemShut
  {NoStop}%
\bibitem [{\citenamefont {Aad}\ \emph {et~al.}(2019)\citenamefont {Aad} \emph
  {et~al.}}]{ATLAS:2019azn}%
  \BibitemOpen
  \bibfield  {author} {\bibinfo {author} {\bibfnamefont {Georges}\ \bibnamefont
  {Aad}} \emph {et~al.} (\bibinfo {collaboration} {ATLAS}),\ }\bibfield
  {title} {\enquote {\bibinfo {title} {{Observation of light-by-light
  scattering in ultraperipheral Pb+Pb collisions with the ATLAS detector}},}\
  }\href {\doibase 10.1103/PhysRevLett.123.052001} {\bibfield  {journal}
  {\bibinfo  {journal} {Phys. Rev. Lett.}\ }\textbf {\bibinfo {volume} {123}},\
  \bibinfo {pages} {052001} (\bibinfo {year} {2019})},\ \Eprint
  {http://arxiv.org/abs/1904.03536} {arXiv:1904.03536 [hep-ex]} \BibitemShut
  {NoStop}%
\bibitem [{\citenamefont {d'Enterria}\ and\ \citenamefont
  {da~Silveira}(2013)}]{dEnterria:2013zqi}%
  \BibitemOpen
  \bibfield  {author} {\bibinfo {author} {\bibfnamefont {David}\ \bibnamefont
  {d'Enterria}}\ and\ \bibinfo {author} {\bibfnamefont {Gustavo~G.}\
  \bibnamefont {da~Silveira}},\ }\bibfield  {title} {\enquote {\bibinfo {title}
  {{Observing light-by-light scattering at the Large Hadron Collider}},}\
  }\href {\doibase 10.1103/PhysRevLett.111.080405} {\bibfield  {journal}
  {\bibinfo  {journal} {Phys. Rev. Lett.}\ }\textbf {\bibinfo {volume} {111}},\
  \bibinfo {pages} {080405} (\bibinfo {year} {2013})},\ \bibinfo {note}
  {[Erratum: Phys.Rev.Lett. 116, 129901 (2016)]},\ \Eprint
  {http://arxiv.org/abs/1305.7142} {arXiv:1305.7142 [hep-ph]} \BibitemShut
  {NoStop}%
\bibitem [{\citenamefont {von Weizsacker}(1934)}]{vonWeizsacker:1934nji}%
  \BibitemOpen
  \bibfield  {author} {\bibinfo {author} {\bibfnamefont {C.~F.}\ \bibnamefont
  {von Weizsacker}},\ }\bibfield  {title} {\enquote {\bibinfo {title}
  {{Radiation emitted in collisions of very fast electrons}},}\ }\href
  {\doibase 10.1007/BF01333110} {\bibfield  {journal} {\bibinfo  {journal} {Z.
  Phys.}\ }\textbf {\bibinfo {volume} {88}},\ \bibinfo {pages} {612--625}
  (\bibinfo {year} {1934})}\BibitemShut {NoStop}%
\bibitem [{\citenamefont {Williams}(1934)}]{Williams:1934ad}%
  \BibitemOpen
  \bibfield  {author} {\bibinfo {author} {\bibfnamefont {E.~J.}\ \bibnamefont
  {Williams}},\ }\bibfield  {title} {\enquote {\bibinfo {title} {{Nature of the
  high-energy particles of penetrating radiation and status of ionization and
  radiation formulae}},}\ }\href {\doibase 10.1103/PhysRev.45.729} {\bibfield
  {journal} {\bibinfo  {journal} {Phys. Rev.}\ }\textbf {\bibinfo {volume}
  {45}},\ \bibinfo {pages} {729--730} (\bibinfo {year} {1934})}\BibitemShut
  {NoStop}%
\bibitem [{\citenamefont {Wang}\ \emph {et~al.}(2022)\citenamefont {Wang},
  \citenamefont {Brandenburg}, \citenamefont {Ruan}, \citenamefont {Shao},
  \citenamefont {Xu}, \citenamefont {Yang},\ and\ \citenamefont
  {Zha}}]{Wang:2022ihj}%
  \BibitemOpen
  \bibfield  {author} {\bibinfo {author} {\bibfnamefont {Xiaofeng}\
  \bibnamefont {Wang}}, \bibinfo {author} {\bibfnamefont {James~Daniel}\
  \bibnamefont {Brandenburg}}, \bibinfo {author} {\bibfnamefont {Lijuan}\
  \bibnamefont {Ruan}}, \bibinfo {author} {\bibfnamefont {Fenglan}\
  \bibnamefont {Shao}}, \bibinfo {author} {\bibfnamefont {Zhangbu}\
  \bibnamefont {Xu}}, \bibinfo {author} {\bibfnamefont {Chi}\ \bibnamefont
  {Yang}}, \ and\ \bibinfo {author} {\bibfnamefont {Wangmei}\ \bibnamefont
  {Zha}},\ }\bibfield  {title} {\enquote {\bibinfo {title} {{Energy Dependence
  of the Breit-Wheeler process in Heavy-Ion Collisions and its Application to
  Nuclear Charge Radius Measurements}},}\ }\href@noop {} {\  (\bibinfo {year}
  {2022})},\ \Eprint {http://arxiv.org/abs/2207.05595} {arXiv:2207.05595
  [nucl-th]} \BibitemShut {NoStop}%
\bibitem [{\citenamefont {Heisenberg}\ and\ \citenamefont
  {Euler}(1936)}]{Heisenberg:1936nmg}%
  \BibitemOpen
  \bibfield  {author} {\bibinfo {author} {\bibfnamefont {W.}~\bibnamefont
  {Heisenberg}}\ and\ \bibinfo {author} {\bibfnamefont {H.}~\bibnamefont
  {Euler}},\ }\bibfield  {title} {\enquote {\bibinfo {title} {{Consequences of
  Dirac's theory of positrons}},}\ }\href {\doibase 10.1007/BF01343663}
  {\bibfield  {journal} {\bibinfo  {journal} {Z. Phys.}\ }\textbf {\bibinfo
  {volume} {98}},\ \bibinfo {pages} {714--732} (\bibinfo {year} {1936})},\
  \Eprint {http://arxiv.org/abs/physics/0605038} {arXiv:physics/0605038}
  \BibitemShut {NoStop}%
\bibitem [{\citenamefont {Adler}\ \emph {et~al.}(2001)\citenamefont {Adler},
  \citenamefont {Denisov}, \citenamefont {Garcia}, \citenamefont {Murray},
  \citenamefont {Strobele},\ and\ \citenamefont {White}}]{Adler:2000bd}%
  \BibitemOpen
  \bibfield  {author} {\bibinfo {author} {\bibfnamefont {Clemens}\ \bibnamefont
  {Adler}}, \bibinfo {author} {\bibfnamefont {Alexei}\ \bibnamefont {Denisov}},
  \bibinfo {author} {\bibfnamefont {Edmundo}\ \bibnamefont {Garcia}}, \bibinfo
  {author} {\bibfnamefont {Michael~J.}\ \bibnamefont {Murray}}, \bibinfo
  {author} {\bibfnamefont {Herbert}\ \bibnamefont {Strobele}}, \ and\ \bibinfo
  {author} {\bibfnamefont {Sebastian~N.}\ \bibnamefont {White}},\ }\bibfield
  {title} {\enquote {\bibinfo {title} {{The RHIC zero degree calorimeter}},}\
  }\href {\doibase 10.1016/S0168-9002(01)00627-1} {\bibfield  {journal}
  {\bibinfo  {journal} {Nucl. Instrum. Meth. A}\ }\textbf {\bibinfo {volume}
  {470}},\ \bibinfo {pages} {488--499} (\bibinfo {year} {2001})},\ \Eprint
  {http://arxiv.org/abs/nucl-ex/0008005} {arXiv:nucl-ex/0008005} \BibitemShut
  {NoStop}%
\bibitem [{\citenamefont {Bertulani}\ \emph {et~al.}(2005)\citenamefont
  {Bertulani}, \citenamefont {Klein},\ and\ \citenamefont
  {Nystrand}}]{Bertulani:2005ru}%
  \BibitemOpen
  \bibfield  {author} {\bibinfo {author} {\bibfnamefont {Carlos~A.}\
  \bibnamefont {Bertulani}}, \bibinfo {author} {\bibfnamefont {Spencer~R.}\
  \bibnamefont {Klein}}, \ and\ \bibinfo {author} {\bibfnamefont {Joakim}\
  \bibnamefont {Nystrand}},\ }\bibfield  {title} {\enquote {\bibinfo {title}
  {{Physics of ultra-peripheral nuclear collisions}},}\ }\href {\doibase
  10.1146/annurev.nucl.55.090704.151526} {\bibfield  {journal} {\bibinfo
  {journal} {Ann. Rev. Nucl. Part. Sci.}\ }\textbf {\bibinfo {volume} {55}},\
  \bibinfo {pages} {271--310} (\bibinfo {year} {2005})},\ \Eprint
  {http://arxiv.org/abs/nucl-ex/0502005} {arXiv:nucl-ex/0502005} \BibitemShut
  {NoStop}%
\bibitem [{\citenamefont {Baltz}\ \emph {et~al.}(2009)\citenamefont {Baltz},
  \citenamefont {Gorbunov}, \citenamefont {Klein},\ and\ \citenamefont
  {Nystrand}}]{Baltz:2009jk}%
  \BibitemOpen
  \bibfield  {author} {\bibinfo {author} {\bibfnamefont {Anthony~J.}\
  \bibnamefont {Baltz}}, \bibinfo {author} {\bibfnamefont {Yuri}\ \bibnamefont
  {Gorbunov}}, \bibinfo {author} {\bibfnamefont {Spencer~R.}\ \bibnamefont
  {Klein}}, \ and\ \bibinfo {author} {\bibfnamefont {Joakim}\ \bibnamefont
  {Nystrand}},\ }\bibfield  {title} {\enquote {\bibinfo {title} {{Two-Photon
  Interactions with Nuclear Breakup in Relativistic Heavy Ion Collisions}},}\
  }\href {\doibase 10.1103/PhysRevC.80.044902} {\bibfield  {journal} {\bibinfo
  {journal} {Phys. Rev. C}\ }\textbf {\bibinfo {volume} {80}},\ \bibinfo
  {pages} {044902} (\bibinfo {year} {2009})},\ \Eprint
  {http://arxiv.org/abs/0907.1214} {arXiv:0907.1214 [nucl-ex]} \BibitemShut
  {NoStop}%
\bibitem [{\citenamefont {Brodsky}\ and\ \citenamefont
  {Lepage}(1981)}]{Brodsky:1981rp}%
  \BibitemOpen
  \bibfield  {author} {\bibinfo {author} {\bibfnamefont {Stanley~J.}\
  \bibnamefont {Brodsky}}\ and\ \bibinfo {author} {\bibfnamefont {G.~Peter}\
  \bibnamefont {Lepage}},\ }\bibfield  {title} {\enquote {\bibinfo {title}
  {{Large Angle Two Photon Exclusive Channels in Quantum Chromodynamics}},}\
  }\href {\doibase 10.1103/PhysRevD.24.1808} {\bibfield  {journal} {\bibinfo
  {journal} {Phys. Rev. D}\ }\textbf {\bibinfo {volume} {24}},\ \bibinfo
  {pages} {1808} (\bibinfo {year} {1981})}\BibitemShut {NoStop}%
\bibitem [{\citenamefont {Zha}\ \emph {et~al.}(2020)\citenamefont {Zha},
  \citenamefont {Brandenburg}, \citenamefont {Tang},\ and\ \citenamefont
  {Xu}}]{Zha:2018tlq}%
  \BibitemOpen
  \bibfield  {author} {\bibinfo {author} {\bibfnamefont {Wangmei}\ \bibnamefont
  {Zha}}, \bibinfo {author} {\bibfnamefont {James~Daniel}\ \bibnamefont
  {Brandenburg}}, \bibinfo {author} {\bibfnamefont {Zebo}\ \bibnamefont
  {Tang}}, \ and\ \bibinfo {author} {\bibfnamefont {Zhangbu}\ \bibnamefont
  {Xu}},\ }\bibfield  {title} {\enquote {\bibinfo {title} {{Initial
  transverse-momentum broadening of Breit-Wheeler process in relativistic
  heavy-ion collisions}},}\ }\href {\doibase 10.1016/j.physletb.2019.135089}
  {\bibfield  {journal} {\bibinfo  {journal} {Phys. Lett. B}\ }\textbf
  {\bibinfo {volume} {800}},\ \bibinfo {pages} {135089} (\bibinfo {year}
  {2020})},\ \Eprint {http://arxiv.org/abs/1812.02820} {arXiv:1812.02820
  [nucl-th]} \BibitemShut {NoStop}%
\bibitem [{\citenamefont {Li}\ \emph {et~al.}(2020)\citenamefont {Li},
  \citenamefont {Zhou},\ and\ \citenamefont {Zhou}}]{Li:2019sin}%
  \BibitemOpen
  \bibfield  {author} {\bibinfo {author} {\bibfnamefont {Cong}\ \bibnamefont
  {Li}}, \bibinfo {author} {\bibfnamefont {Jian}\ \bibnamefont {Zhou}}, \ and\
  \bibinfo {author} {\bibfnamefont {Ya-Jin}\ \bibnamefont {Zhou}},\ }\bibfield
  {title} {\enquote {\bibinfo {title} {{Impact parameter dependence of the
  azimuthal asymmetry in lepton pair production in heavy ion collisions}},}\
  }\href {\doibase 10.1103/PhysRevD.101.034015} {\bibfield  {journal} {\bibinfo
   {journal} {Phys. Rev. D}\ }\textbf {\bibinfo {volume} {101}},\ \bibinfo
  {pages} {034015} (\bibinfo {year} {2020})},\ \Eprint
  {http://arxiv.org/abs/1911.00237} {arXiv:1911.00237 [hep-ph]} \BibitemShut
  {NoStop}%
\bibitem [{\citenamefont {Reece}\ and\ \citenamefont
  {Wang}(2009)}]{Reece:2009un}%
  \BibitemOpen
  \bibfield  {author} {\bibinfo {author} {\bibfnamefont {Matthew}\ \bibnamefont
  {Reece}}\ and\ \bibinfo {author} {\bibfnamefont {Lian-Tao}\ \bibnamefont
  {Wang}},\ }\bibfield  {title} {\enquote {\bibinfo {title} {{Searching for the
  light dark gauge boson in GeV-scale experiments}},}\ }\href {\doibase
  10.1088/1126-6708/2009/07/051} {\bibfield  {journal} {\bibinfo  {journal}
  {JHEP}\ }\textbf {\bibinfo {volume} {07}},\ \bibinfo {pages} {051} (\bibinfo
  {year} {2009})},\ \Eprint {http://arxiv.org/abs/0904.1743} {arXiv:0904.1743
  [hep-ph]} \BibitemShut {NoStop}%
\bibitem [{\citenamefont {Fayet}(2017)}]{fayet2017light}%
  \BibitemOpen
  \bibfield  {author} {\bibinfo {author} {\bibfnamefont {Pierre}\ \bibnamefont
  {Fayet}},\ }\bibfield  {title} {\enquote {\bibinfo {title} {{The light $U$
  boson as the mediator of a new force, coupled to a combination of $Q,B,L$ and
  dark matter}},}\ }\href@noop {} {\bibfield  {journal} {\bibinfo  {journal}
  {The European Physical Journal C}\ }\textbf {\bibinfo {volume} {77}},\
  \bibinfo {pages} {1--12} (\bibinfo {year} {2017})}\BibitemShut {NoStop}%
\bibitem [{\citenamefont {Aaron}\ \emph {et~al.}(2010)\citenamefont {Aaron}
  \emph {et~al.}}]{H1:2009cml}%
  \BibitemOpen
  \bibfield  {author} {\bibinfo {author} {\bibfnamefont {F.~D.}\ \bibnamefont
  {Aaron}} \emph {et~al.} (\bibinfo {collaboration} {H1}),\ }\bibfield  {title}
  {\enquote {\bibinfo {title} {{Diffractive Electroproduction of rho and phi
  Mesons at HERA}},}\ }\href {\doibase 10.1007/JHEP05(2010)032} {\bibfield
  {journal} {\bibinfo  {journal} {JHEP}\ }\textbf {\bibinfo {volume} {05}},\
  \bibinfo {pages} {032} (\bibinfo {year} {2010})},\ \Eprint
  {http://arxiv.org/abs/0910.5831} {arXiv:0910.5831 [hep-ex]} \BibitemShut
  {NoStop}%
\bibitem [{\citenamefont {STAR}()}]{BUR}%
  \BibitemOpen
  \bibfield  {author} {\bibinfo {author} {\bibnamefont {STAR}},\ }\href@noop {}
  {\enquote {\bibinfo {title} {{STAR Public Note SN0793 - The STAR Beam Use
  Request for Run-23-25}},}\ }\bibinfo {howpublished}
  {\url{https://drupal.star.bnl.gov/STAR/starnotes/public/SN0793}}\BibitemShut
  {NoStop}%
\bibitem [{\citenamefont {Klein}\ \emph {et~al.}(2020)\citenamefont {Klein},
  \citenamefont {Mueller}, \citenamefont {Xiao},\ and\ \citenamefont
  {Yuan}}]{Klein:2020jom}%
  \BibitemOpen
  \bibfield  {author} {\bibinfo {author} {\bibfnamefont {Spencer}\ \bibnamefont
  {Klein}}, \bibinfo {author} {\bibfnamefont {A.~H.}\ \bibnamefont {Mueller}},
  \bibinfo {author} {\bibfnamefont {Bo-Wen}\ \bibnamefont {Xiao}}, \ and\
  \bibinfo {author} {\bibfnamefont {Feng}\ \bibnamefont {Yuan}},\ }\bibfield
  {title} {\enquote {\bibinfo {title} {Lepton pair production through two
  photon process in heavy ion collisions},}\ }\href {\doibase
  10.1103/PhysRevD.102.094013} {\bibfield  {journal} {\bibinfo  {journal}
  {Phys. Rev. D}\ }\textbf {\bibinfo {volume} {102}},\ \bibinfo {pages}
  {094013} (\bibinfo {year} {2020})}\BibitemShut {NoStop}%
\end{thebibliography}%

\end{document}